\documentclass[twocolumn,showpacs,prb,amsmath]{revtex4}
\usepackage{epsfig,graphicx}

\def\be{\begin{equation}}
\def\ee{\end{equation}}
\def\e#1{\label{#1}\end{equation}}
\def\bea{\begin{eqnarray}}
\def\eea{\end{eqnarray}}
\def\ea#1{\label{#1}\end{eqnarray}}
\def\rqn#1{(\ref{#1})}
\def\bes#1{\begin{subequations}\label{#1}}
\def\ese{\end{subequations}}

\begin{document}
\title{Bell inequality violation versus entanglement in presence of
local decoherence}
\author{A.\ G.\ Kofman}
\author{A.\ N.\ Korotkov}
\affiliation{Department of Electrical Engineering, University of
California, Riverside, California 92521}
\date{\today}

\begin{abstract}
We analyze the effect of local decoherence of two qubits on their
entanglement and the Bell inequality violation. Decoherence is
described by Kraus operators, which take into account dephasing and
energy relaxation at an arbitrary temperature.
 We show that in the experiments with superconducting phase qubits
the survival time for entanglement should be much longer than for
the Bell inequality violation.
    \end{abstract}
    \maketitle

Entanglement of separated systems is a genuine quantum effect and an
essential resource in quantum information processing. \cite{nie00}
Experimentally, a convincing evidence of a two-qubit entanglement is
a violation of the Bell inequality \cite{Bell} in its
Clauser-Horne-Shimony-Holt\cite{CHSH} (CHSH) form. However, only for
pure states the entanglement always \cite{cap73} results in a
violation of the Bell inequality. In contrast, some mixed entangled
two-qubit states (as we will see, most of them) do not violate the
Bell inequality, \cite{wer89} though they may still exhibit
nonlocality in other ways. \cite{pop94} Distinction between
entanglement and Bell-inequality violation, in its relevance to
experiments with superconducting phase qubits,\cite{ste06} is the
subject of our paper.

   The two-qubit entanglement is usually characterized by
the concurrence \cite{woo98} $C$ or by the entanglement of
formation,\cite{Bennett-96} which is a monotonous function
\cite{woo98} of $C$. Non-entangled states have $C=0$, while $C=1$
corresponds to maximally entangled states. There is a
straightforward way\cite{woo98} to calculate $C$ for any two-qubit
density matrix $\rho$. The Bell inequality in the CHSH form
\cite{CHSH} is $|S|\leq 2$, where
$S=E(\vec{a},\vec{b})-E(\vec{a},\vec{b}')+ E(\vec{a}',\vec{b})+
E(\vec{a}',\vec{b}')$ and $E(\vec{a},\vec{b})$ is the correlator of
results ($\pm 1$) for measurement of two qubits (pseudospins) along
directions $\vec{a}$ and $\vec{b}$. This inequality should be
satisfied by any local hidden-variable theory, while in quantum
mechanics it is violated up to $|S|=2\sqrt{2}$ for maximally
entangled (e.g., spin-zero) states. Mixed states produce smaller
violation (if any), and there is a straightforward way \cite{hor95}
to calculate the maximum value $S_+$ of $|S|$ for any two-qubit
density matrix.

    For states with a given concurrence $C$, there is an exact bound
\cite{ver02} for $S_+$:  $2\sqrt{2}C\leq S_+\leq 2\sqrt{1+C^2}$ (we
consider only $S_+>2$), so that the Bell inequality violation is
guaranteed if $C>1/\sqrt{2}$. For any pure state the upper bound is
reached: $S_+=2\sqrt{1+C^2}$, so that non-zero entanglement always
leads to $S_+>2$.
 The distinction between entanglement
and Bell inequality violation has been well studied for so-called
Werner states \cite{wer89} which have the form $\rho=f
\rho_s+(1-f)\rho_{\rm mix}$, where $\rho_s$ denotes the maximally
entangled (singlet) state, and $\rho_{\rm mix}={\bf 1}/4$ is the
density matrix of the completely mixed state. The Werner state is
entangled for\cite{wer89} $f>1/3$, while it violates the Bell
inequality only when\cite{hor95} $f>1/\sqrt{2}$.

    The Werner states, however, are not relevant to most of
experiments (including those with superconducting phase qubits
\cite{ste06}), in which an initially pure state becomes mixed due to
decoherence (Werner states are produced due to so-called
depolarizing channel\cite{nie00}).
 Recently a number of authors have analyzed effects of qubit
decoherence on the Bell inequality violation
\cite{sam03,beenak03,jak04,sli05,jam06} and entanglement.
\cite{Loss03,tyu04,tyu06,tol05,ged06,san06,Nori06}
    Best-studied models of decoherence in this context are
pure dephasing \cite{sam03,beenak03,sli05,tol05,ged06,Nori06} and
zero-temperature energy relaxation, \cite{jak04,jam06,tyu04,san06}
while there are also papers considering a combination of these
mechanisms, \cite{Loss03,tyu06} high-temperature energy relaxation,
\cite{jak04} and non-local decoherence. \cite{jak04,sli05,Nori06}
 In particular, for the case of pure dephasing it has been shown
\cite{tol05,tyu06} that the concurrence $C$ decays as a product of
decoherence factors for the two qubits, and therefore a state
remains entangled for arbitrarily long time; moreover, the
calculation of $S_+$ shows\cite{sam03,beenak03} that the Bell
inequality is always violated also. For the case of zero-temperature
energy relaxation it has been shown that entanglement can still last
forever\cite{tyu04,san06,jam06} (depending on the initial state),
while a finite survival time has been obtained \cite{jam06} for the
Bell inequality violation.

    In this paper we consider a two-qubit state decoherence due to
general (Markovian) local decoherence of each qubit (including
dephasing and energy relaxation at a finite temperature) and assume
absence of any other evolution. For this model we compare for how
long an initial state remains entangled ($C>0$), and for how long it
can violate the Bell inequality ($S_+>2$). In particular, we show
that for typical (best) present-day parameters for phase
qubits\cite{ste06} these durations differ by $\sim 8$ times.

    Before analyzing this problem let us discuss which fraction of
the entangled two-qubit states violate the Bell inequality. This
question is well-posed only if we introduce a particular metric
(distance) and corresponding measure (volume) in the 15-dimensional
space of density matrices. Various metrics are possible; let us
choose the Hilbert-Schmidt metric, \cite{nie00,zyc01} for which the
geometry in the space of states is Euclidean.
 Then random states $\rho$ with the uniform probability distribution
can be generated as \cite{zyc01} $\rho=A^\dagger A/{\rm
tr}(A^\dagger A)$, where $A$ is a $4\times 4$ matrix, all elements
of which are independent Gaussian complex variables with the same
variance and zero mean.
 Using this method, we performed Monte-Carlo simulation, generating
$10^9$ random states and checking if they are entangled
\cite{note1,per96,san98} and if they violate the Bell
inequality.\cite{hor95}
 In this way we confirmed that 75.76\% of all states are entangled
\cite{slater-07} and found that only 0.822\% of all states violate
the Bell inequality. Therefore, only a small fraction, 1.085\% of
entangled states violate the Bell inequality.

    Now let us discuss the effect of decoherence.
    For one qubit it can be described by the Bloch
equations \cite{coh92} (we use the basis of the ground state
$|0\rangle$ and excited state $|1\rangle$) and characterized by the
energy relaxation time $T_1$, dephasing time $T_2$ ($T_2\leq 2
T_1$), and the Boltzmann factor $h=\exp (-\Delta /\theta )$, where
$\Delta$ is the energy separation of the states and $\theta$ is the
temperature. The usual solution of the Bloch equations can be
translated into the language of time-dependent superoperator ${\cal
L}$ for the one-qubit density matrix $\rho$, so that $\rho(t)={\cal
L}[\rho(0)]=\sum_{i=1}^4 K_i\rho(0)K_i^\dagger$, where four Kraus
operators $K_i$ can be chosen as

 \bea
K_1=\left(\begin{array}{cc}
0&0\\
\sqrt{g}&0
\end{array}\right), \,\,\,\,\,\,
 K_2=\left(\begin{array}{cc}
\sqrt{1-g}&0\\0&\lambda/\sqrt{1-g}
\end{array}\right),\nonumber\\
K_3=\left(\begin{array}{cc} 0&0\\0
&\sqrt{1-hg-\frac{\lambda^2}{1-g}}
\end{array}\right), \,\,\,\,
K_4=\left(\begin{array}{cc}
 0&\sqrt{hg}\\0&0
\end{array}\right) ,
 \ea{5.8-m}
where $g=[1-\exp(-t/T_1)]/(1+h)$, $\lambda=\exp (-t/T_2)$, and in
our notation $|1\rangle=(1,0)^T$, $|0\rangle=(0,1)^T$. It is easy to
check that the term under the square root in $K_3$ is always
non-negative and equals 0 (for $t>0$) only if $T_2=2T_1$ and
$\theta=0$. Notice that choice of the Kraus operators $K_i$ is not
unique (though limited to the unitary freedom of quantum
operations\cite{nie00}) and, for instance, the Kraus operators
presented in Ref.\ \onlinecite{nie00} for the special cases of
depolarizing channel ($T_1=T_2$, $\theta =\infty$) and energy
relaxation ($T_2=2T_1$) differ from Eq.\ (\ref{5.8-m}).

    In general, decoherence of two qubits is described by many
parameters (out of 240 parameters describing a general quantum
operation only 15 parameters describe unitary evolution). We choose
a relatively simple but physically relevant model when the
decoherence is dominated by local decoherence of each qubit.
(Non-local decoherence would be physically impossible in the case of
large distance between the qubits.) The model now involves six
parameters: $T_1^{a,b}$, $T_2^{a,b}$, and
$h_{a,b}=\exp(-\Delta_{a,b}/\theta_{a,b})$, where subscripts (or
superscripts) $a$ and $b$ denote qubits, and the evolution is
described by the tensor-product superoperator ${\cal L}={\cal
L}_a\otimes{\cal L}_b$ (which is completely positive because of
complete positivity of ${\cal L}_{a,b}$). This superoperator
contains 16 terms: $\rho(t)={\cal L}[\rho(0)]
=\sum_{i,j=1}^4K_{ij}\rho(0)K_{ij}^\dagger$, $K_{ij} = K_i^a\otimes
K_j^b$, where operators $K_i^{a,b}$ are given by Eq.\ (\ref{5.8-m})
for each qubit.

As an initial state we consider an ``odd'' pure state
 \be
|\Psi\rangle=\cos \beta  \, |10\rangle+ e^{i\alpha}\sin \beta \,
|01\rangle
 \e{2.5}
($0<\beta<\pi/2$), which is relevant for experiments with the phase
qubits. \cite{ste06} Since the parameter $\alpha$ corresponds to
$z$-rotation of one of the qubits, while decoherence as well as
values of $C$ and $S_+$ are insensitive to such rotation, all
results of our model have either trivial or no dependence on
$\alpha$. The evolution of the state (\ref{2.5}) due to local
decoherence  ${\cal L}$ can be calculated analytically, and at time
$t$ the non-vanishing elements of the two-qubit density matrix
$\rho$ are
 \bea
&& \rho_{11}(t)=(1-g_a) h_b g_b \cos^2 \beta  + h_ag_a(1-g_b)\sin^2
\beta  ,
    \nonumber \\
&& \rho_{22}(t)=(1-g_a) (1-h_b g_b) \cos^2 \beta + h_ag_a g_b \sin^2
\beta  ,
    \nonumber\\
&& \rho_{33}(t) = g_a h_bg_b \cos^2 \beta
+(1-h_ag_a)(1-g_b)\sin^2\beta ,
    \nonumber \\
&& \rho_{44}(t) = g_a (1-h_bg_b) \cos^2 \beta  +(1-h_ag_a) g_b
\sin^2 \beta ,
    \nonumber\\
&& \rho_{32}(t)= \rho_{23}^*(t)= \exp(-t/T_2^a-t/T_2^b) e^{i\alpha}
(\sin 2\beta)/2 , \qquad
 \ea{2}
where $g_{a,b}$ are defined below Eq.\ (\ref{5.8-m}), and $\rho_{ij}$
subscripts $i,j=1,2,3,4$ correspond to the basis $\{|11\rangle,
|10\rangle, |01\rangle,|00\rangle\}$. These equations become very
simple at zero temperature because then $h_a=h_b=0$. Notice that the
dephasing times $T_2^{a,b}$ enter Eqs. \rqn{2} only through the
combination $1/T_2^a+1/T_2^b$ (this is not so for a general initial
state), so that the two-qubit dephasing can be characterized by one
parameter $T_2\equiv 2/(1/T_2^a+1/T_2^b)$.

   For the state \rqn{2} the concurrence is  \cite{tyu06,jak04}
 \be
C=2\max\{0,|\rho_{23}|-\sqrt{\rho_{11}\rho_{44}}\},
 \e{3}
and the Bell inequality parameter $S_+$ is \cite{hor95,jam06}
 \be
S_+=2\max\{2\sqrt{2}|\rho_{23}|,
\sqrt{4|\rho_{23}|^2+(1-2\rho_{11}-2\rho_{44})^2}\} ,
 \e{5}
while for the initial state $C=\sin 2\beta >0$ and
$S_+=2\sqrt{1+C^2}>2$. Notice that the first and second terms in
Eq.\ (\ref{5}) correspond to the ``horizontal'' and ``vertical''
measurement configurations, using the terminology of Ref.\
\onlinecite{Kofman-07}. Equations (\ref{2}), (\ref{3}), and
(\ref{5}) are all we need to analyze entanglement and Bell
inequality violation.

    Notice that for a pure
dephasing ($T_1^a=T_1^b=\infty$) we have $\rho_{11}=\rho_{44}=0$,
and therefore
 \be
C=\exp (-2t/T_2) \sin 2\beta , \,\,\, S_+=2\sqrt{1+C^2}.
 \e{4}
In this case at any $t$ the state remains entangled
\cite{tol05,tyu06} and violates the Bell inequality.
\cite{sam03,beenak03} (It also remains within the class of states
producing maximal Bell inequality violation for a given
concurrence.\cite{ver02}) In the case when both dephasing and energy
relaxation are present but temperature is zero,
$\theta_a=\theta_b=0$, the concurrence $C$ is still given by Eq.\
(\ref{4}) and lasts forever; \cite{san06,jam06} however $S_+$ does
not satisfy Eq.\ (\ref{4}) and, most importantly, the Bell
inequality is no longer violated after a finite time.\cite{jam06}
Finally, in presence of energy relaxation at non-zero temperature
(at least for one qubit) the entanglement also vanishes after a
finite time, as seen from Eq.\ (\ref{3}), in which
$\lim_{t\rightarrow\infty} \rho_{11}\rho_{44} \neq 0$.

    Let us consider in more detail the case when both dephasing
and energy relaxation are present, but temperature is zero and
$T_1^a=T_1^b\equiv T_1$. Then Eq.\ (\ref{5}) for $S_+$ becomes very
simple since $\rho_{11}=0$ and $\rho_{44}=1-\exp (-t/T_1)$. The time
dependence $S_+(t)$ consists of three regions: at small $t$ it is
always determined by the second term \cite{note-beta} in Eq.\
(\ref{5}), then after some time $t_1$ the first term becomes
dominating, while after a later time $t_2$ the second term becomes
dominating again. Notice that in the second region
$S_+=4\sqrt{2}|\rho_{23}|=2\sqrt{2}C$, so such state provides
minimal $S_+$ for a given concurrence $C$. \cite{ver02,note} The
time $\tau_B$ after which the Bell inequality is no longer violated
[$S_+(\tau_B)=2$] falls either into the first or second region,
because $S_+(t_2)<2$ [it is interesting to note that in the third
region $S_+(t)$ passes through a minimum and then increases up to
$S_+ \rightarrow 2$ at $t\rightarrow \infty$]. The time $\tau_B$ can
be easily calculated if $S_+(t_1)>2$, so that $\tau_B$ falls into
the second region and therefore
    \begin{equation}
    \tau_B= (T_2/2) \ln (\sqrt{2}\sin 2\beta).
    \label{tau-B-an}\end{equation}
This case is realized when pure dephasing is relatively weak:
$T_1/T_2\le\ln(\sqrt{2} \sin 2\beta)/[2\ln(4-2\sqrt{2})]$; since
$T_1/T_2\geq 1/2$, it also requires $\sin 2\beta \ge 2\sqrt{2}-2$.
[For $T_1/T_2=1/2$ Eq.\ (\ref{tau-B-an}) has been obtained in Ref.\
\onlinecite{jam06}.] Notice that $\tau_B$ in Eq.\ (\ref{tau-B-an})
corresponds to the condition $C=1/\sqrt{2}$, while in general
$\tau_B$ corresponds to $C\leq 1/\sqrt{2}$ because of the
inequality\cite{ver02} $S_+\geq 2\sqrt{2}C$.

\begin{figure}[tb]
\includegraphics[width=7.8cm]{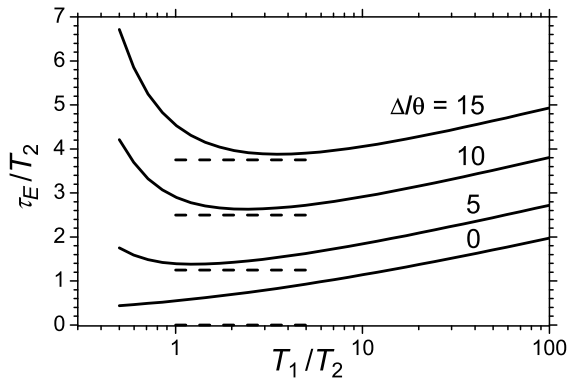}
\caption{The two-qubit entanglement duration $\tau_E$ in units of
the dephasing time $T_2$ for the maximally entangled initial state
($\beta=\pi/4$) and several values of the temperature $\theta$.
Dashes lines correspond to Eq.\ (\ref{tau-E}).
   }
 \label{f1}\end{figure}

Now let us focus on calculating the duration $\tau_E$ of the
entanglement survival, duration $\tau_B$ of the Bell inequality
violation, and their ratio $\tau_E/\tau_B$ at non-zero temperature.
For simplicity we limit ourselves to the case of maximally entangled
initial state ($\beta =\pi /4$), and we also assume equal energy
relaxation, splitting and temperature for both qubits:
$T_1^a=T_1^b\equiv T_1$, $\Delta_a=\Delta_b\equiv \Delta$, and
$\theta_a=\theta_b\equiv \theta$ (we do not need to assume equal
dephasing, since it can be characterized by only one parameter
$T_2$).
   As follows from Eq.\ \rqn{3}, the entanglement duration $\tau_E$ can be
calculated numerically using the equation
 $|\rho_{23}|=\sqrt{\rho_{11}\rho_{44}}$.
 Figure \ref{f1} shows $\tau_E$ (normalized by $T_2$) as a function of
the ratio $T_1/T_2$ for several values of the normalized inverse
temperature $\Delta/\theta$. As we see, in a typical experimental
regime \cite{ste06} when $\Delta /\theta \sim 10$, the ratio
$\tau_E/T_2$ does not depend much on $T_1/T_2$ when $T_1$ is larger
but comparable to $T_2$ (which is also typical experimentally). In
other words, $\tau_E$ is approximately proportional to $T_2$, and in
this regime $\tau_E$ also has crudely inverse dependence on
temperature [see Eq.\ (\ref{tau-E}) below].

    Analytical formulas for $\tau_E$ can be easily obtained in the
limiting cases. In absence of pure dephasing ($T_1/T_2=1/2$) and low
temperature ($\theta \ll \Delta$) we find
 $\tau_E/T_2 \approx \Delta/2\theta -\ln(2\sqrt{2}+2)/2 \approx \Delta/2\theta -0.79$,
while at high temperature ($\theta \gg \Delta$) we have
 $\tau_E/T_2 \approx \ln(\sqrt{2}+1)/2 \approx 0.44$.
In the case of strong dephasing ($T_1/T_2\gg 1$) we find (neglecting
some corrections) $\tau_E/T_2\simeq\Delta/(4\theta
)+\ln(T_1/T_2)/2$.

    However, these asymptotic formulas are not very relevant to a
typical experimental situation with phase qubits,\cite{ste06} in
which $T_1\agt T_2$. As another way to approximate $\tau_E$ we have
chosen the value at the minimum of the curves in Fig.\ \ref{f1};
this minimum occurs at the ratios $T_1/T_2$ somewhat close to the
experimental values, and the result is naturally not much sensitive
to $T_1/T_2$ in a significantly broad range. For sufficiently small
temperatures ($\Delta/\theta >2$) we have obtained approximation
$(\tau_E/T_2)_{\rm min} \approx \Delta/4\theta
+\ln(3^{3/4}/2)\approx\Delta/4\theta +0.13$ and found that the
minimum occurs at
$T_1/T_2 \approx (\tau_E/T_2)_{\rm min}/\ln3$. So, as the crudest
approximation in the experimentally-relevant regime ($\theta /
\Delta \sim 10^{-1}$, $T_1/T_2\agt 1$), the two-qubit entanglement
lasts for (see dashed lines in Fig.\ \ref{f1})
    \begin{equation}
\tau_E \simeq T_2 \Delta /4\theta .
    \label{tau-E}    \end{equation}

\begin{figure}[tb]
\includegraphics[width=7.9cm]{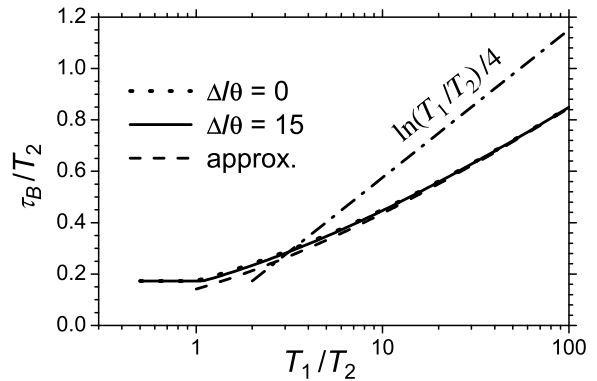}
\caption{The duration $\tau_B$ of the  Bell inequality violation
(assuming $\beta =\pi/4$) for $\Delta/\theta =15$ (solid line) and
$\Delta/\theta =0$ (dotted line).
 The dashed line:
 $\tau_B/T_2=\ln[T_1/(4\tau_B)]/4$.
 }
 \label{f3}\end{figure}

    The duration $\tau_B$ of the Bell inequality violation is
calculated using Eq.\ \rqn{5} as $S_+(\tau_B)=2$. Solid and dotted
lines in Fig.\ \ref{f3} show numerical results for $\tau_B$ (in units
of $T_2$) as a function of the ratio $T_1/T_2$ for low and high
temperatures: $\Delta/\theta =15$ and 0. The curves are almost
indistinguishable, that means that $\tau_B$ is practically
independent of the temperature for fixed $T_1$ and $T_2$. Notice that
each curve consists of a constant (horizontal) part and an increasing
part, which correspond to two terms in Eq.\ \rqn{5}. It can be shown
that at zero temperature the horizontal part is realized at
$T_1/T_2\le\ln2/[4\ln(4-2\sqrt{2})]\approx 1.1$, while at high
temperature ($\theta \gg \Delta$) it is realized at $ T_1/T_2\le 1$.
The horizontal part corresponds to the first term in Eq.\ \rqn{5}
dominating at $\tau_B$: $S_+=2\sqrt{2}\exp(-2 t/T_2)$, so at
sufficiently weak pure dephasing we have $\tau_B/T_2=\ln2/4\approx
0.17$ [see also Eq.\ (\ref{tau-B-an})]. In the opposite case of
strong pure dephasing ($T_1/T_2\gg 1$) the duration $\tau_B$ is the
solution of the equation
$\tau_B/T_2=\ln[T_1/(4\tau_B)]/4$ (dashed line in Fig.\ \ref{f3}),
so roughly $\tau_B/T_2\simeq\ln(T_1/T_2)/4$ (dot-dashed line in
Fig.\ \ref{f3}). Combining these results, we get a crude estimate:
    \begin{equation}
    \tau_B \simeq T_2 \max\{0.17, \, 0.25 \ln (T_1/T_2)\}.
    \label{tau-B}    \end{equation}

\begin{figure}[tb]
\includegraphics[width=7.9cm]{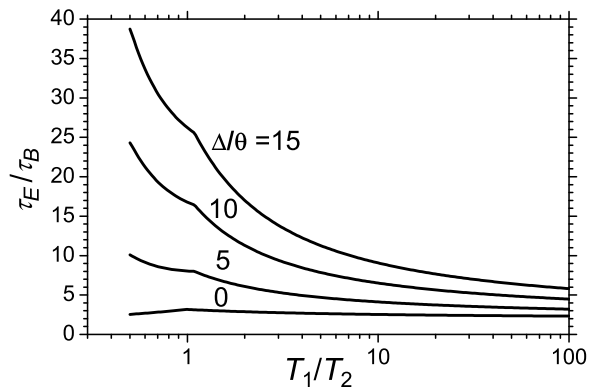}
\caption{The ratio $\tau_E/\tau_B$ for the maximally entangled
initial state and several values of the temperature $\theta$. }
 \label{f4}\end{figure}

Figure \ref{f4} shows the ratio $\tau_E/\tau_B$ of the survival
durations of entanglement and the Bell inequality violation. We see
that the ratio $\tau_E/\tau_B$ increases with the decrease of
temperature and decrease of the pure dephasing contribution, which
are both the desired experimental regimes. (This rule does not work
in the experimentally irrelevant regime $\theta \gg \Delta$ and
$T_1<T_2$.) Notice that the kinks on the curves correspond to the
change of the dominating term in Eq.\ \rqn{5}.  In absence of pure
dephasing ($T_1/T_2=1/2$) the low-temperature result ($\theta \ll
\Delta$) is $\tau_E/\tau_B\approx (2/\ln2)
[\Delta/\theta-\ln(2\sqrt{2}+2)]$, while at $\theta \gg \Delta$ the
ratio is $\tau_E/\tau_B\approx 2\ln(\sqrt{2}+1)/\ln2\approx 2.5$. In
the limit of strong pure dephasing ($T_1/T_2\gg 1$) the asymptotic
result is $\tau_E/\tau_B\approx 2+(\Delta/\theta)/\ln(T_1/T_2)$ (as
we see, $\tau_E
> 2\tau_B$ for any parameters).
   In the experimentally relevant regime when $\theta /\Delta \sim 10^{-1}$ and $T_1/T_2\agt
  1$, the ratio can be obtained from Eqs.\ (\ref{tau-E}) and
(\ref{tau-B}), giving a crude estimate
 $\tau_E/\tau_B \simeq (\Delta
/\theta)\min \{1.5,\, 1/\ln(T_1/T_2)\}$.

   For an experimental estimate let us choose parameters typical for
best present-day experiments with superconducting phase qubits:
\cite{ste06} $\Delta/2\pi\hbar\simeq6$ GHz, $\theta\simeq 50$ mK,
$T_1\simeq 450$ ns, $T_2\simeq300$ ns. Then $\Delta/\theta\simeq 6$,
$T_1/T_2\simeq 1.5$, and we obtain $\tau_E \simeq 470$ ns, $\tau_B
\simeq 60$ ns, and $\tau_E/\tau_B\simeq 7.7$.

    In conclusion, we have found that in the Hilbert-Schmidt metric
only 1.085\% of entangled states violate the Bell inequality, thus
explaining why entanglement can last for a significantly longer time
($\tau_E$) than the Bell inequality violation ($\tau_B$). Using the
technique of Kraus operators, we have considered local decoherence
due to dephasing and energy relaxation at finite temperature, and
for this model calculated $\tau_E$, $\tau_B$, and their ratio
$\tau_E/\tau_B$.
    The work was supported by NSA and DTO under ARO grant W911NF-04-1-0204.


\begin{thebibliography}{99}
\bibitem{note1}
Entanglement is checked by the fast method  based on the sign of the
determinant of the partially transposed state \cite{per96}
$\tilde{\rho}$, using the fact that for an entangled state $\rho$
all eigenvalues of $\tilde{\rho}$ are non-zero, and exactly one of
them is negative. \cite{san98}
\bibitem{note-beta}
Notice that when $T_1^a \neq T_1^b$, the second term in Eq.\
(\ref{5}) is maximized for a non-maximally entangled state, $\beta
\neq \pi/4$, though the benefit is not significant if we need
$S_+\agt 2.2$.
\bibitem{note}
The statement in Ref.\ \protect\onlinecite{ver02} that any mixed
state with $S_+=2\sqrt{2}C>2$ is maximally entangled, is incorrect
(here maximum entanglement means that $C$ cannot be increased by any
two-qubit unitary transformation). As a counterexample, consider the
states $\rho = f |\Psi \rangle \langle \Psi | +(1-f) |00 \rangle
\langle 00 |$, produced from the initial state \protect\rqn{2.5} due
to zero-temperature energy relaxation ($T_2=2T_1$, $\theta=0$,
$f=e^{-t/T_1}$). Any two such states with the same $f$ but different
initial parameter $\beta$ can obviously be connected by a unitary
transformation (involving only the subspace spanned by $|01\rangle $
and $|10\rangle$), while they have different concurrence $C$ given by
Eq.\ (\ref{4}). Finally, as follows from our analysis, there is a
finite range of parameters $f$ and $\beta$, in which
$S_+=2\sqrt{2}C$; in this range the concurrence can still be varied
by unitary transformations varying $\beta$, contradicting the
statement of Ref.\ \protect\onlinecite{ver02}.
\end{thebibliography}
\end{document}